\documentclass[%
 reprint,
superscriptaddress,
 amsmath,amssymb,
 aps,
floatfix,
]{revtex4-2}

\usepackage{siunitx}

\usepackage{placeins}

\usepackage[hidelinks]{hyperref}

\usepackage{graphicx}
\usepackage{dcolumn}
\usepackage{bm}

\begin{document}

\preprint{APS/123-QED}



\title{Picosecond X-ray Imaging of Shockwaves with Non-Rankine-Hugoniot Behavior}

\author{Christopher S. Campbell}
\affiliation{Department of Mechanical Engineering, Texas A\&M University, College Station, Texas 77843, USA}
\affiliation{Los Alamos National Laboratory, Los Alamos, New Mexico 87545, USA}

\author{Mirza Akhter}%
\affiliation{Department of Mechanical Engineering, Texas A\&M University, College Station, Texas 77843, USA}

\author{Samuel Clark}
\author{Kamel Fezzaa}
\affiliation{X-ray Science Division, Advanced Photon Source, Argonne National Laboratory, Argonne, Illinois 60439, USA}

\author{Zhehui Wang}
\affiliation{Los Alamos National Laboratory, Los Alamos, New Mexico 87545, USA}
 \email{zwang@lanl.gov}

\author{David Staack}
\affiliation{Department of Mechanical Engineering, Texas A\&M University, College Station, Texas 77843, USA}
 \email{dstaack@tamu.edu}

\date{\today}

\begin{abstract}

The first-known observation of plasma-induced cavitation bubbles and expanding shockwaves in liquid during plasma initiation timescales reveals deviation from expected Rankine-Hugoniot shock behavior due to coupled shock-cavitation dynamics, imaged using megahertz-framerate picosecond X-ray imaging.
The imaging target features an inexpensive benchtop-scale pulsed plasma device used to generate well-timed spark discharges in ambient liquid heptane at an unprecedented repetition rate ($>$3/min) compared with more commonly used dynamic targets.
These shockwaves are relatively weak (Mach number $\leq$ 1.4) compared with X-ray-imaged shockwaves in prior literature, advancing the resolution and sensitivity limits of this high-speed imaging diagnostic.
Phase contrast imaging (PCI) has facilitated enhanced quantitative analysis of the expanding shocks in this work, via comparison to thermodynamic models and a Fresnel-Kirchhoff diffraction model.

\end{abstract}

\maketitle



In high-speed X-ray science, efficient use of limited beamtime at advanced light source user facilities hinges on the maximum achievable repetition rate of dynamic targets, which typically correlates with the ability to make more nuanced scientific insights.
However, most dynamic processes of interest involve destructible devices \cite{Ram2012,Dattelbaum2020ShockwaveStructures,Sechrest2020}, necessitating constant target reassembly and restricting event rates to only a few events per hour.
It is therefore beneficial to develop dynamic targets which require minimal to no maintenance between events, without compromising phenomena of interest such as high instantaneous power density, high mass density gradients, high pressure and temperature gradients, and supersonic behavior/shockwaves \cite{Im2009,OMalley2014,Magaletti2015ShockNanobubble,Perkins2009}.
In this letter we present such a target, a pulsed plasma device submerged in ambient liquid heptane which can produce expanding shockwaves at a rate exceeding 3 events per minute.
Our analysis, facilitated by the high data rate, reveals a deviation in shockwave behavior from the accepted Rankine-Hugoniot model, which we attribute to the interaction between the shock front and the expanding cavitation bubble associated with the submerged pulsed plasma event—a phenomenon
To the best our knowledge, this letter contains the first experimental evidence of this of this effect.


The shock front images presented herein also constitute one of the weakest shock fronts ever imaged using high-speed radiographic techniques (Ma $\approx$ 1.2).
While a sufficiently strong shock would be easily visible using less sensitive imaging techniques due to its relatively high mass density ratio, the fact that such a weak shock front is still observable in this work highlights the superior sensitivity of this implementation of PCI to very subtle dynamic phenomena, while still revealing the limits of current techniques and the path forward for the next generation of ultrafast X-ray imaging.
The high repetition rate ($>$3 events/minute), low cost ($<$US\$100k), and portability of this imaging target make it quite attractive to those fields interested in similar dynamic events (e.g. ICF, dynamic compression, shock physics), but which rely on apparatuses which are either immovable or have a prohibitively slow event repetition rate \cite{Chu2023_PLX,Gao2022,Nora2015GigabarLaser}.



The pulsed power device and high-voltage circuit utilized in this study to generate submerged spark discharge is akin to those in our previous work \cite{Campbell2021,Akhter2021}, which consists of two electrodes between which a well-timed submerged spark discharge occurs (see Section SM.II).
The event of interest dissipates approximately 100mJ of nanosecond-timescale plasma processes (light, sound, chemistry, shockwaves) in the target over 100ns, implying an instantaneous power of roughly 1MW.
Assuming an approximate discharge cross section of $5 \si{\micro\meter}$ during peak current across a gap of 0.5mm, we estimate a peak surface power density of $4 \si{\tera\watt/\centi\meter^2}$ and a peak energy density per unit mass of $15 \si{\giga\joule/\kg}$, within two orders of magnitude (albeit at a lower instantaneous power) of the $1 \si{\tera\joule/kg}$ implied by recent hotspot energy and mass results from the National Ignition Facility \cite{Zylstra2022}.

\begin{figure}[!t] 
    \centering
    \includegraphics[width=\linewidth]{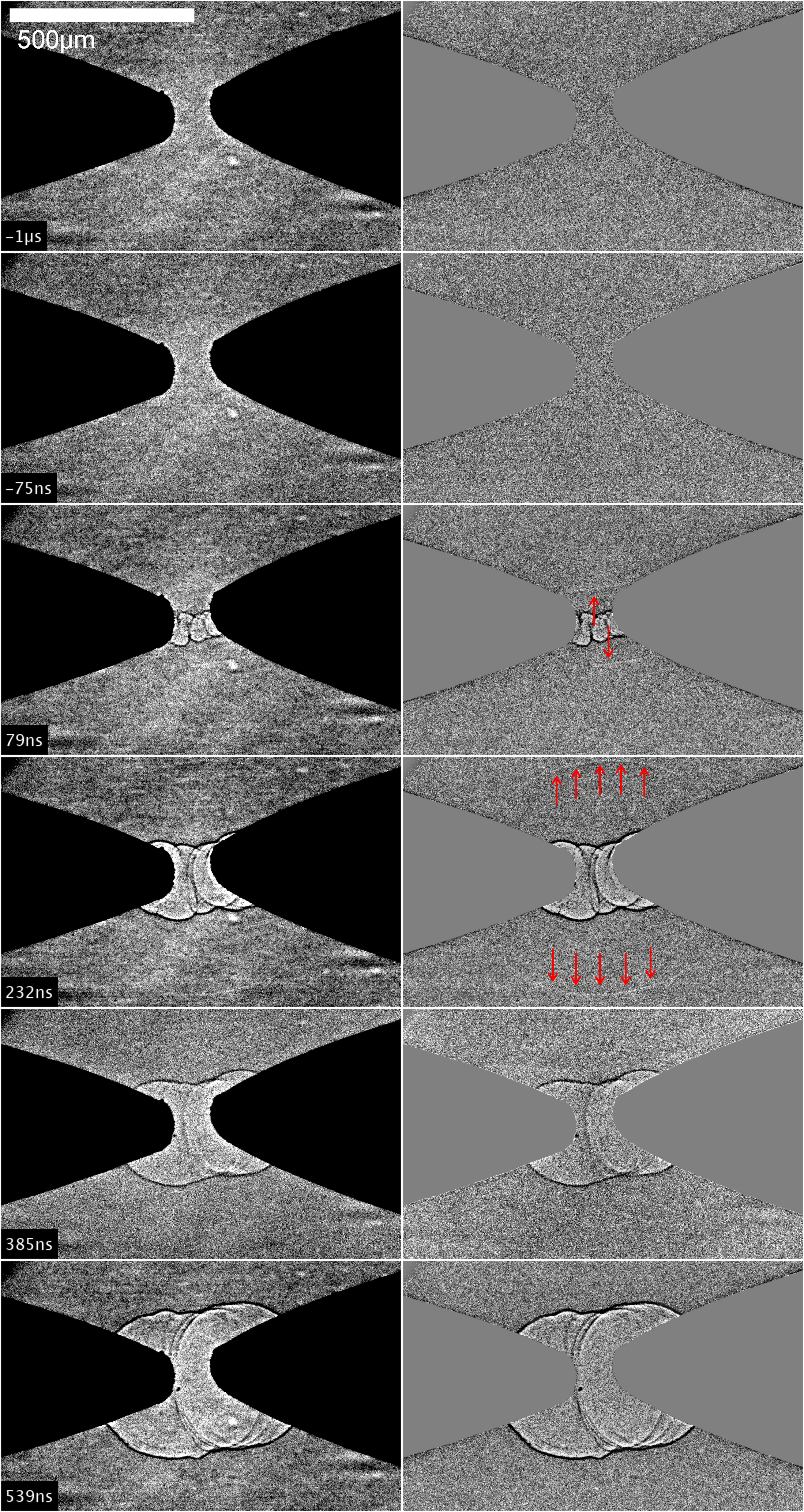} 
    \caption{
    Selected frames from a single spark discharge event in heptane for which the plasma-induced shock is visible, with timestamps measured relative to spark initiation.
    Left and right columns show contrast-enhanced raw images and corresponding background-subtracted frames respectively.
    Note the location of the shock front visible at t=79ns and t=232ns, denoted by red arrows.
    }
    \label{fig:singleevent}
\end{figure}

\begin{figure}[!t]
    \centering
    \includegraphics[width=\linewidth]{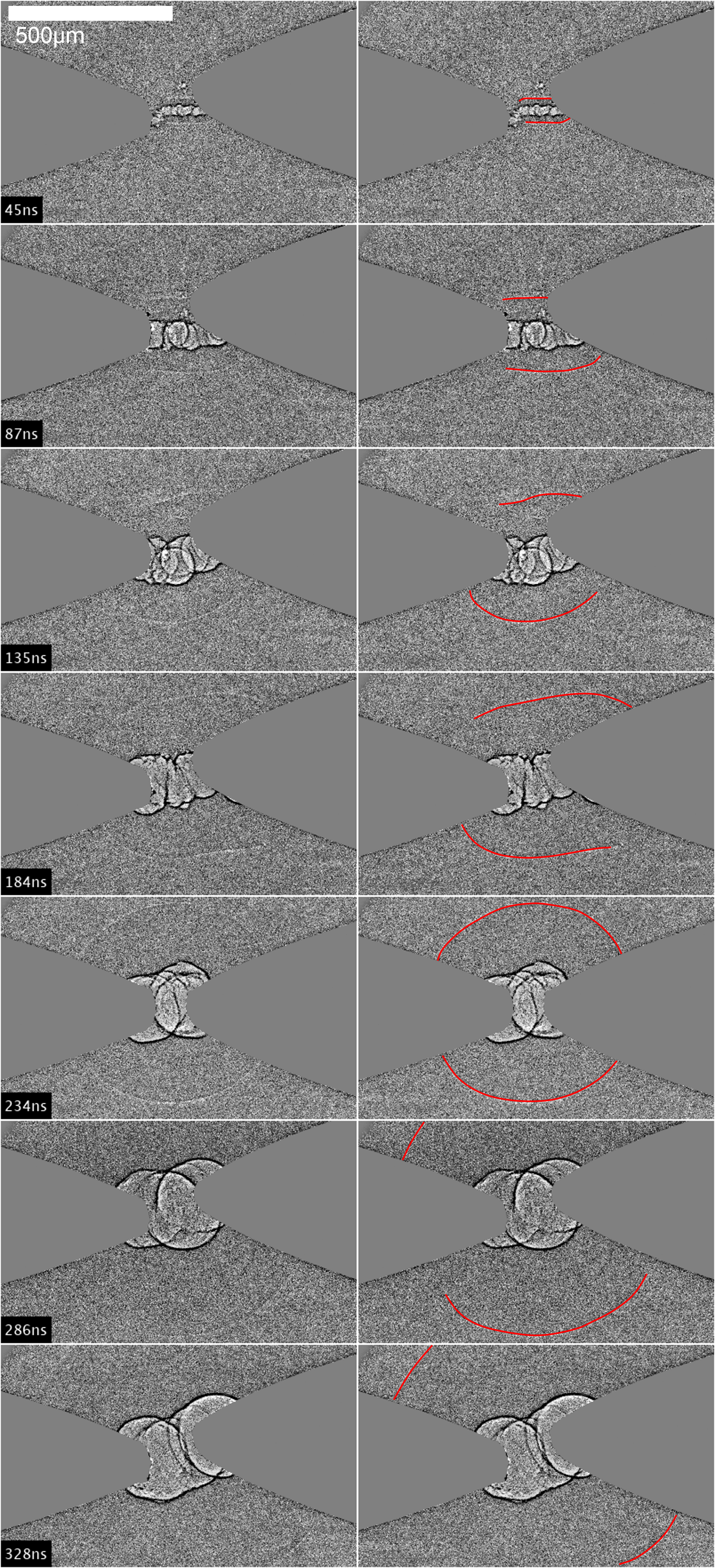}
    \caption{
    Frames from selected PCI heptane spark events in which a shock front was visible, sorted by frame time relative to spark initiation, which the right column.
    Each frame is duplicated across both columns, the right column includes annotations (red) which indicate the contour of the shock.
    \vspace{0.25in}
    }
    \label{fig:sortedframes}
\end{figure}

\begin{figure}[!t]
    \centering
    \includegraphics[width=\linewidth]{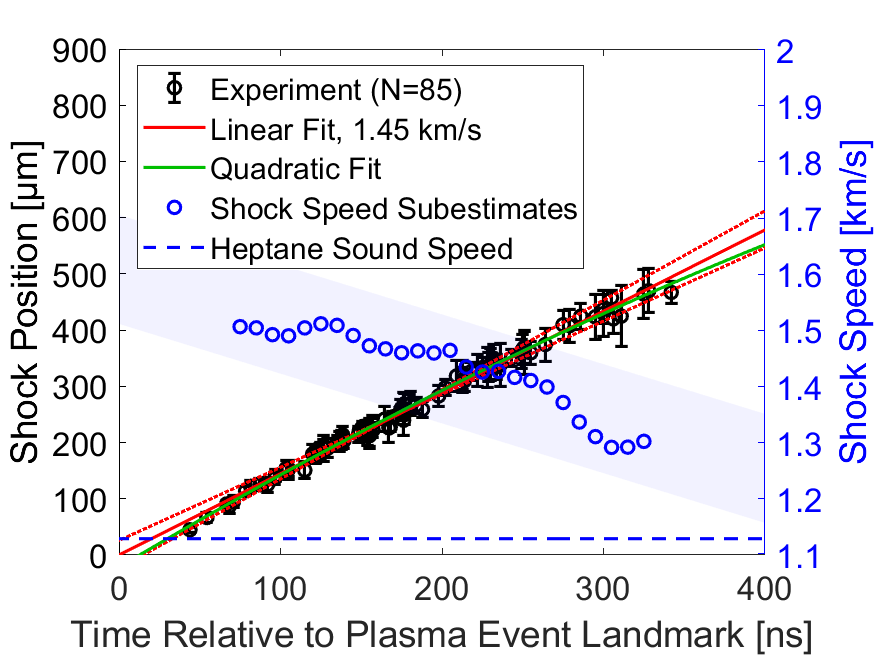}
    \caption{
    Plot of cylindrical shock positions/radii, relative to the axis of the plasma, compiling data from eighty-five different PCI frames for which the shock was visible (Figure \ref{fig:sortedframes}).
    Shock position measurement was performed manually for each of the 85 frames in which a shock was visible.
    The twenty measurements for each frame were then used to determine uncertainty (two standard deviations away from the average).
    The solid line (red) shows the least-squares linear fit to these points (1.45 km/s), with the two red dotted lines assume a shock speed 90\% and 110\% of that linear fit to roughly illustrate inherent shock speed uncertainty.
    Blue circles indicate shock speed subestimates calculated by fitting to subsamples of the full set of shock positions.
    The green curve shows a quadratic fit to the position vs. time data.
    }
    \label{fig:shockspeed}
\end{figure}

\begin{figure}[!t]
    \centering
        \includegraphics[width=\linewidth]{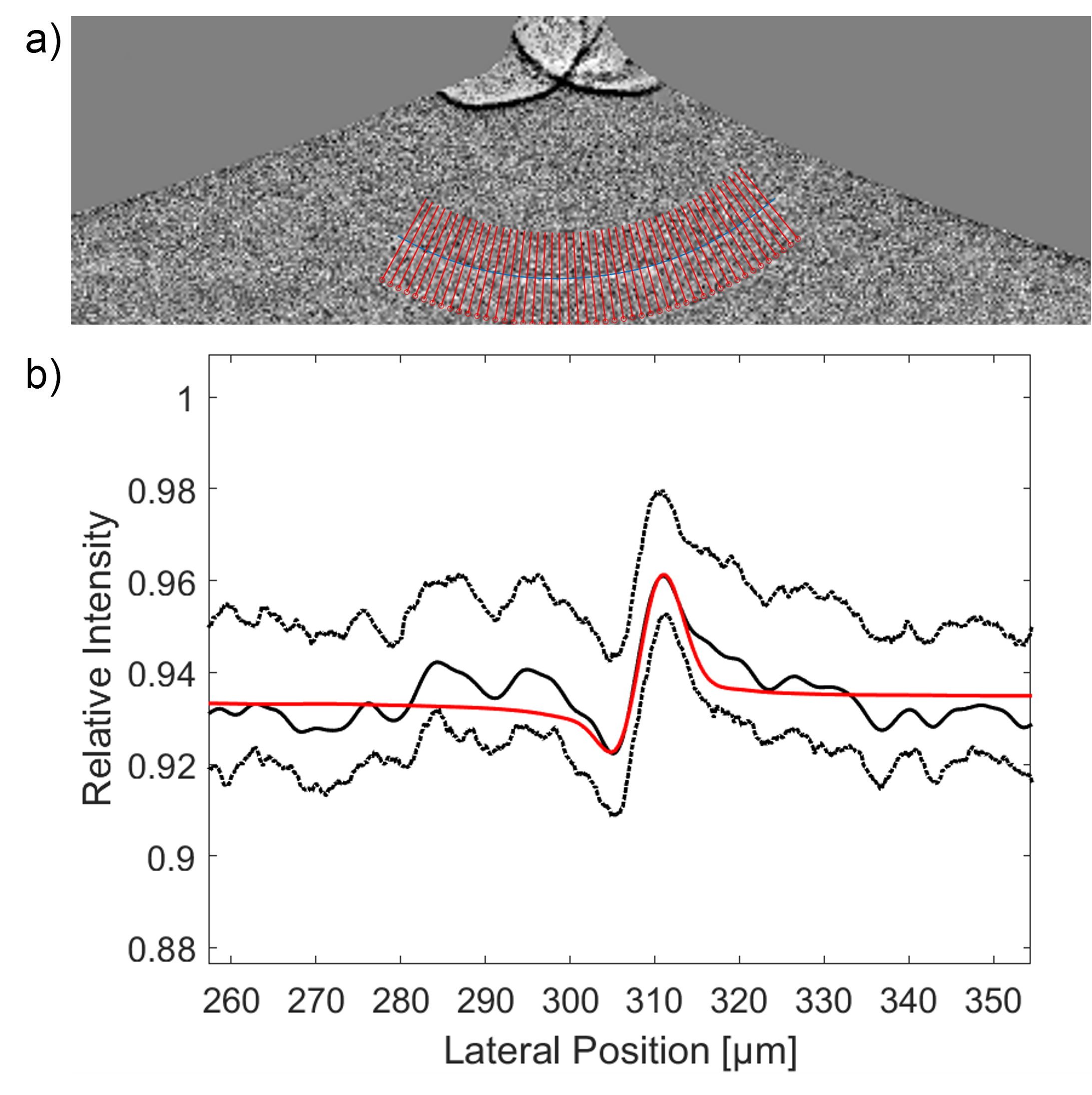}
    \caption{
    Illustration of a cutline extraction algorithm which uses a spline fit to convert the two-dimensional X-ray frame of the shock front (a) into a one-dimensional plot (b).
    The image analyzed here corresponds with the fifth image from Figure \ref{fig:sortedframes} (t=234ns).
    The black solid line in represents the average cutline (dotted lines show upper and lower quartiles), and the red line shows the best simulated shock front PCI profile with a post-shock density of $\rho_2 = 0.799^{+0.116}_{-0.057} \si{g/cc}$ ($\rho_2/\rho_1 = 1.176^{+0.171}_{-0.084}$).
    }
    \label{fig:comparetomodel}
\end{figure}

This target was taken to the Argonne National Laboratory's Advanced Photon Source (APS) for ultrafast X-ray phase-contrast imaging (PCI) experiments.
The X-ray imaging method for this work used a 128-frame Shimadzu HPV-X2 camera ($3 \si{\micro\meter}$/pixel) to image a LYSO scintillator in line with the synchrotron source and imaging target (see Section SM.II).
The 6.5MHz imaging framerate was enabled by the APS's 24-singlet standard operating mode \cite{APSparams}.
Figure \ref{fig:singleevent} shows selected PCI frames from a single spark discharge event, while Figure \ref{fig:sortedframes} compiles frames from multiple similar events, sorted by frame time relative to plasma initiation.

Figure \ref{fig:shockspeed} shows an estimated shock speed of $1.45 \pm 0.13 \si{km/s}$, corresponding to a Mach number of $1.28 \pm 0.13$ in ambient heptane ($v_\text{sound} = 1.129 \si{km/s}$).
The shock images' transverse profile aligns with the expected PCI for a step discontinuity in density.
A slight negative concavity in the data implies that the shock speed decreases with time, consistent with the time-dependent shock speed from linear fits to subsamples within 100ns of a given time.
Although Taylor–von Neumann–Sedov blast wave theory might seem applicable (where the time-dependent shock position $R(t)$ is proportional to $t^{0.4}$ for spherically expanding shocks \cite{Taylor1950,Grun1991} and to $t^{0.5}$ for cylindrically expanding shocks \cite{Lin1954}), the imaged plasma-induced shock front contradicts the theory's two main assumptions: instantaneous energy input and negligible ambient pressure ($p_\text{post-shock} \gg p_\text{ambient}$).
By the time the shock becomes visible to this diagnostic (earliest measured shock image at 45ns after initiation), the post-shock pressure has decreased drastically, resulting in a near-linear position vs. time relation.
By 45ns after initiation, the significantly reduced post-shock pressure results in nearly linear position-time relation, aligning with O'Malley's experimental \cite{OMalley2014} and Magaletti's computational results \cite{Magaletti2015ShockNanobubble} for expanding shocks in liquids.
Therefore, we deemed a phenomenological quadratic fit (Figure \ref{fig:shockspeed}, green curve) to be appropriate as it minimizes assumptions while capturing this negative concavity.


The X-ray absorption contrast for such a weak shock is quite low; in our imaging target, absorption contrast for the shock is $\sim$0.02\% (6mm path length through heptane, $100\si{\micro\meter}$ shocked region, 20\% density increase), well below the detectability limit for this experiment.
However, the edge enhancement properties of PCI cause this shock to be visible above background noise as localized maxima and minima in brightness, with the maxima occurring on the pre-shock side of the discontinuity.
This type of diffraction is the essence of PCI and is governed by the Fresnel–Kirchhoff integral \cite{MITocw_fresnel}, modified for translationally symmetric geometries in Equation \ref{eq:fresnel} to ease computation:
\begin{flalign}
    g_\text{out}(x',y') &= \frac{e^{2\pi i z/\lambda}}{i \lambda z} \iint g_\text{in}(x,y) e^{\frac{i \pi}{\lambda z} ((x'-x)^2 + (y'-y)^2)   }dx dy \\
    \label{eq:fresnel} &= \frac{e^{2\pi i z/\lambda}}{\sqrt{i \lambda z}} \int g_\text{in} (x) e^{\frac{i \pi}{\lambda z} (x'-x)^2} dx
\end{flalign}
where $g_\text{in}$ and $g_\text{out}$ represent the complex-valued electric field at the target and the imaging plane respectively, $\lambda$ is the X-ray wavelength being used ($\lambda=$ $\sim$25keV in our case), and $z$ is the target-to-detector distance ($z = 46 \si{\centi\meter}$).
All information about the target is contained in $g_\text{in}$ as complex-valued index of refraction data, taken from \cite{Henke_1993}.
See Section SM.III for more detail on Equation \ref{eq:fresnel} derivation.
This model can now be fit to experiment (Figure \ref{fig:comparetomodel}, analyzing the fifth frame from Figure \ref{fig:sortedframes}), constituting a measurement technique to estimate post-shock density.

\begin{figure}[!t]
    \centering
    \includegraphics[width=1\linewidth]{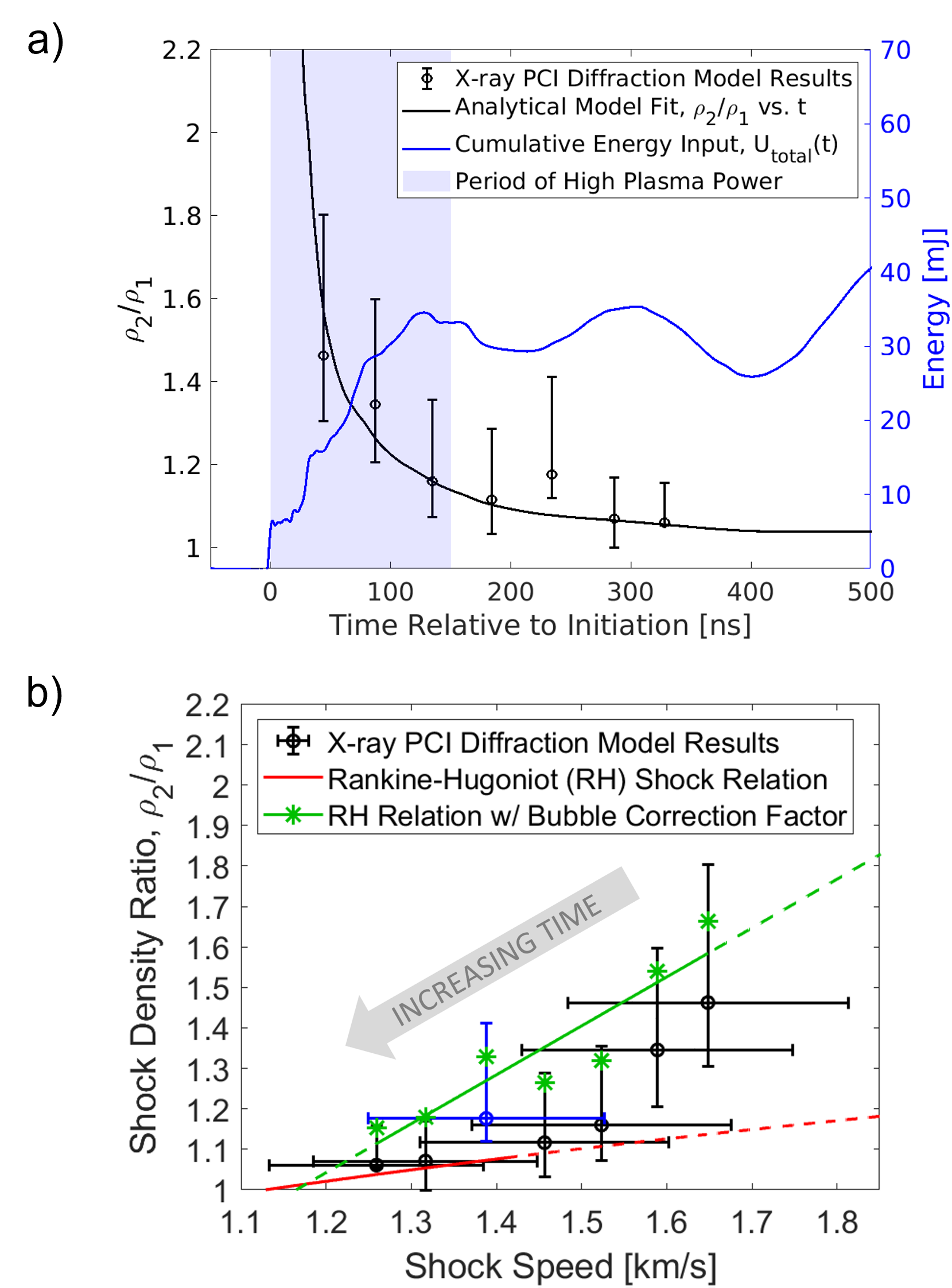}
    \caption{
    Summary plots of the PCI-enabled density measurements performed in this work.
    The upper plot (a) shows how shock density ratio decays with time, along with the analytical fit from Equation \ref{eq_rhovstime}.
    The lower plot (b) depicts Hugoniot states in the $\rho_2$--$v_\text{shock}$ space, showing how values estimated from this work's PCI data and diffraction model (black) compare to normal shock thermodynamic relations in heptane both with (green) and without (red) the cavitation bubble compression effect.
    The blue datapoint corresponds with the particular X-ray diffraction model fit result from Figure \ref{fig:comparetomodel}.
    }
    \label{fig:densityVSspeed}
\end{figure}

This density measurement method was used to analyze a subset of seven PCI frames from multiple events featuring a visible shock image.
Figure \ref{fig:densityVSspeed}a uses knowledge of relative frame delay to show time-dependent behavior of the shock density ratio $\rho_2/\rho_1$; the observed decay in $\rho_2/\rho_1$ over time arises purely from computational analysis of the PCI images, and is consistent with expanding shock front behavior.
We identify the cumulative excess thermal energy over time in the post-shock region as $U_\text{shock}(t)$, assumed to be proportional to measured cumulative electrical energy input $U_\text{total}(t)$, also included in Figure \ref{fig:densityVSspeed}a.
By further assuming that the shock expands roughly spherically, energy density in the post shock region bounded between the shock front and the cavitation bubble surface  can be expressed as $\Delta u(t) = U_\text{shock}(t)/\big(\rho_2(t) \cdot \frac{4}{3}\pi\cdot(R_\text{shock}(t)^3 - R_\text{bubble}(t)^3)\big)$, where $\Delta u(t) = u_2(t) - u_1$ is the excess specific energy above ambient.
After incorporating a power law for heptane's $\rho_2/\rho_1$ vs. $\Delta u$ relation \cite{Span2003}, we arrive at:
\begin{flalign}
    \frac{\rho_2(t)}{\rho_1} = A\cdot\Bigg(\frac{U_\text{shock}(t)}{\rho_2(t) \cdot \frac{4}{3}\pi\cdot(R_\text{shock}(t)^3 - R_\text{bubble}(t)^3)}\Bigg)^B+1 \label{eq_rhovstime}
\end{flalign}
\noindent where $A = 2.05 \times 10^{-3} \si{\kilo\gram^B\joule^{-B}}$ and $B=0.415$ are the power law parameters.
$U_\text{shock}(t)$, $R_\text{shock}(t)$ and $R_\text{bubble}(t)$ are determined from electrical diagnostics and imaging data.
By implicitly solving for $\rho_2(t)$, we can now include the analytical model from Equation \ref{eq_rhovstime} in Figure \ref{fig:densityVSspeed}a, which agrees well with the PCI model results.
This thermodynamic model is related to the phenomenological fit to $\rho_2/\rho_1$ vs. $t$ data from diamond shock compression PCI experiments by Schropp at SLAC's LCLS \cite{Schropp2015}.
A notable result of this physics-based analysis is measurement of $U_\text{shock}(t)$ as a fraction of $U_\text{total}(t)$; in this case that fraction (found by fitting to experiment) is $\alpha = U_\text{shock}(t)/U_\text{total}(t) = 2.8\%$.






Alongside the X-ray diffraction model method for measuring $\rho_2/\rho_1$, it is possible to relate $\rho_2$ and $v_\text{shock}$ by solving the system of Rankine–Hugoniot thermodynamic shock relations in heptane (see Section SM.I for derivation), resulting in a unique relationship between $\rho_2$ and $v_\text{shock}$ (Figure \ref{fig:densityVSspeed}b, red curve).
However, measured values for $\rho_2$ and $v_\text{shock}$ (from the diffraction model and quadratic fit to Figure \ref{fig:shockspeed}) tend to reside above this red curve, suggesting a higher-density post-shock region than what is implied by this thermodynamic model.
We attribute this to the fact that the thermodynamic model ignores the cavitation bubble which follows the expanding shock, clearly visible in Figures \ref{fig:singleevent} and \ref{fig:sortedframes}.
This bubble interface is generated at a relatively small radius ($<10 \si{\micro\meter}$) and high energy density, and expands outward with a large amount of inertia, compressing the post-shock region to a higher density.
This is shown with green asterisks in Figure \ref{fig:densityVSspeed}b, generated by assuming that this compression only affects the post-shock density without affecting the speed of the shock; this is implemented by multiplying the density ratio from the original RH relation by the compression ratio $1/\big(1-(\frac{R_\text{bubble}}{R_\text{shock}})^2\big)$.
Although this correction factor still does not result in a model that sufficiently matches experiment, it represents the maximum post-shock density $\rho_2$ which could be justified by the data.
The original thermodynamic model (red curve) represents the minimum bound on $\rho_2$, since it completely neglects this compression effect.
In reality, the compression-induced higher $\rho_2$ should increase the shock speed, an effect not fully explored here.
Full investigation of this effect should include a multiphysics model which couples together the dynamics of the cavitation bubble (Rayleigh-Plesset), the liquid post-shock region (Navier–Stokes), and the shock jump conditions (Rankine-Hugoniot).

Similarity of this work to the submerged exploding wire PCI experiments by Yanuka \cite{Yanuka2018} at the European Synchrotron Radiation Facility (ESRF) allows us to directly compare these two different implementations of PCI and of cylindrically expanding shock events, of which the most dramatic difference is peak instantaneous power and total event energy; for Yanuka, these values were approximately 1GW and 300J, respectively (recall 1MW and 100mJ for this work).
The fact that the shock front images presented in this work are still visible with such a small event energy demonstrates the superior sensitivity and utility of PCI as a diagnostic for analyzing lower-energy shocks.



In summary, we present robust and repeatable observation of weak shocks in liquid heptane and their interaction with the plasma-induced cavitation bubble at timescales previously obscured by optical emission.
This achievement constitutes a feat in imaging sensitivity and resolution, and advances an underexplored approach for in-depth quantitative analysis of subtle phenomena using PCI.
The enhanced capability of the featured imaging target to generate large datasets at a rate as high as 1Hz is particularly promising for machine learning applications; the eighty-five shock fronts cataloged in Figure \ref{fig:shockspeed} could be used in the future to train a deep learning model for rapid analysis and computation of shock parameters, since it rapidly and stochastically queries a wide range of parameters (imaging delay time, shock shape, breakdown voltage, etc.).
This advantage would also still hold for plasma-induced shock events in other media of interest.
This avenue also shows promise for potentially observing higher-order shock structure and nonuniformity (see Section SM.VI).
Future endeavors will focus on extending the PCI sensitivity threshold to shock imaging while continuing to develop a quantitative analysis toolkit, for both shock imaging and general PCI applications.


\begin{acknowledgements}
The authors would like to acknowledge the support of the U.S. DOE NNSA. Los Alamos National Laboratory is managed by Triad National Security, LLC for the U.S. Department of Energy's NNSA.
This document has been approved for release under No. LA-UR-23-22175.
This research used resources of the Advanced Photon Source, a U.S. Department of Energy (DOE) Office of Science user facility operated for the DOE Office of Science by Argonne National Laboratory under Contract No. DE-AC02-06CH11357.
\end{acknowledgements}

\bibliography{mybib}
\bibliographystyle{ieeetr}

\end{document}